\documentclass[proof]{WileyASNA-v1}

\articletype{Article Type}%

\received{1 October 2020}
\revised{12 October 2020}
\accepted{25 October 2020}

\begin{document}

\title{Constraints on Bose-Einstein condensate stars as neutron stars models from new observational data}

\author[1]{Adriel E. Rodr\'iguez Concepci\'on}
\author[2,1]{Gretel Quintero Angulo}

\authormark{Rodr\'iguez Concepci\'on \& Quintero Angulo}

\address[1]{\orgdiv{Facultad de F\'isica}, \orgname{Universidad de La Habana}, \orgaddress{\state{La Habana}, \country{Cuba}}}

\address[2]{\orgdiv{Institut f\"{u}r Theoretische Physik}, \orgname{Heinrich-Heine-Universit\"{a}t}, \orgaddress{\state{D\"{u}sseldorf}, \country{Germany}}}

\corres{Gretel Quintero Angulo, Institut f\"{u}r Theoretische Physik, Heinrich-Heine-Universit\"{a}t D\"{u}sseldorf,\\ Universit\"{a}tsstr.\,1, 40225 D\"{u}sseldorf, Germany\\  \email{Gretel.Quintero.Angulo@hhu.de}}

\abstract{We evaluate the feasibility of Bose-Einstein condensate stars (BECS) as models for the interior of neutron stars. BECS are compact objects composed of bosons, formed through the spin-parallel pairing of neutrons. Here, we utilize the astronomical data from GW170817, XMMU J173203.3-344518 (the lightest neutron star known), and a novel lower limit on neutron star core heat capacity to scrutinize the compatibility of BECS with these recent observations of neutron stars. Our specific focus is to constrain the values of the scattering length $a$, a parameter determining the strength of particle interactions in the model. Our analysis suggests that if the stars involved in GW170817 were BECSs, the scattering length of their constituent bosons should fall within the $4$ to $10$ fm range. Additionally, at a scattering length of $a\sim 3.1-4$ fm, stars with mass and radius characteristics akin to XMMU J173203.3-344518 are identified. Moreover, we find that the heat capacity depends of the mass and temperature of BECS, and surpasses the established lower bound for neutron star cores when $a>2-5$ fm. In summary, our results endorse BECS models with $a\sim 4$ fm, providing neutron star observables in robust agreement with diverse observations and contributing to the ongoing understanding of neutron star interiors.}


\maketitle

\section{Introduction}\label{sec1}

The inner regions of neutron stars (NS) are one of the most exciting and intriguing scenarios of current physics research. From exotic condensates to several states of deconfined quarks, many kinds of matter and phases have been conjectured to exist in NS interiors, although, so far, the knowledge extracted from observations is insufficient to decide in favor of one of those multiple options. Among the possibilities, is that of superfluid NS cores \cite{Migdal}. Observational evidence associated with the theoretical adjustment of the cooling data of Cassiopeia A has come to reinforce this long-lasting supposition \cite{Shternin2011,Page:2011yz,PageSC}. Considering NS's typical inner conditions, paired neutrons should be in an intermediate situation between the Bardeen-Copper-Schrieffer and the Bose-Einstein condensate (BEC) limits, prevailing the former description \cite{Gruber2014}. However, assuming the pairs in the BEC limit led, some years ago, to Bose-Einstein condensate stars (BECS) models \cite{chavanis2012bose}.

A BECS is a compact object composed of a condensed gas of interacting bosons where gravity is balanced by the pressure that comes from the interactions. In the context of NS, these bosons are formed by the spin-parallel pairing of neutrons, so that the bosons are neutral, have a mass twice the neutron mass, and spin one \cite{PageSC}. The characteristic densities of BECS of paired neutrons ($\sim10^{14}-10^{16}$ g/cm$^3$) are close to those of the cores of  NS. Since in a NS, the core contributes the majority of the mass and size, naked BECSs\footnote{In this context naked means no crust no atmosphere} may properly approach NS in the limiting case their core is mostly composed of tightly paired neutrons \cite{Gruber2014}, as long their macroscopic properties are compatible with observations. 

The mass and size of BECS are mainly controlled by the mass and dispersion length of the bosons \cite{chavanis2012bose}, but may be also influenced by the temperature of the star and its magnetic field \cite{Latifah,Gruber2014,quintero2019self,Lis2}. In this paper, we aim to evaluate BECS as NS models in the light of some late astronomical observations, in particular, those regarding the detection of an ultralight NS (so far associated with strange stars models) \cite{Doroshenko2022}, the measurement of the tidal deformability of neutron stars \cite{Abbott2019}, and the estimation of a lower bound for the specific heat of compact objects \cite{Cumming2017}.

In Section \ref{sec2} we briefly review the equation of state of BECS, while Section \ref{sec3} is devoted to the formalism for computing the macroscopic properties. Finally, in Section \ref{sec4} we constrain the range of values of the scattering length $a$ and the star temperature so that the resulting BECSs fit the observational data. Section \ref{sec5} encloses our conclusions. In this paper we use natural units in which $\hbar=k_B=c=\frac{1}{4\pi\epsilon_0}=1$.

\section{Equation of state for Bose-Einstein condensate stars}
\label{sec2}

As stated in the introduction, the BECSs we consider are compact stars composed of interacting neutral vector bosons with mass $m=2m_n$, being $m_n$ the neutron mass. In a first approximation, the interaction between the bosons can be considered as a two-body contact interaction \cite{Latifah}. The strength of this interaction is determined by the parameter $u_0 = \frac{4\pi a}{m}$, where $a \sim 1 - 10$ fm is the bosons scattering length, and the mean field thermodynamic potential for the interacting bosons can be written as 
\begin{equation}\label{Omega1}
\Omega_T(T,\mu) = \Omega(T,\mu) + \frac{1}{2} u_0 N^2_T.
\end{equation}
The mean field approximation holds as far as the quantum ﬂuctuations are negligible, but this is to expect for a dense system as the BECS at the relatively low temperatures inside the star\footnote{Compare the typical NS temperatures $T\sim10^{6}$K with the low temperature limit of the boson gas $T<<m\sim10^{13}$K.} \cite{Lis2}.
Despite its simplicity, the assumption of a two-body contact interaction allows us to obtain thermodynamic consistent EoS for BECS whose mass-radius curves are consistent with NS observations \cite{Latifah}. 

The first term of Eq.~(\ref{Omega1}) is the thermodynamic potential of the free non-relativistic bosons \cite{quintero2021}
\begin{equation}\label{Omega2}
	\Omega(T,\mu) = - \dfrac{2s+1}{\beta^{5/2}} \left(\dfrac{m}{2\pi}\right)^{\frac{3}{2}} P_{5/2} (e^{\beta\mu'}),
\end{equation}
where $\beta=1/T$ is the inverse temperature, $\mu$ is the chemical potential, $s=1$ is the particle spin, $\mu^{\prime} = \mu - m$ and $P_{l}(z)$ is the polylogarithmic function of order $l$. The second term stands for the energy of the interactions, and depends on the total particle density $N_T = N_{gs} + N(T,m)$, being $N_{gs}$ the density of particles in the condensate and $N(T,\mu)$  the density of particles in the excited states \cite{quintero2021}
\begin{equation}
	\label{Nfree}
	N(\beta,\mu) = - \left(\dfrac{\partial \Omega}{\partial \mu}\right)_T = (2s+1)\left( \dfrac{m}{2\pi\beta}\right)^{\frac{3}{2}}P_{3/2} (e^{\beta\mu'}).
\end{equation}

It is worth noticing that, for the scope of this work, we will consider non-magnetized BECS, so that the non-relativistic description of the interacting boson gas will suffice. On the contrary, in the case of magnetized BECS, a relativistic EoS is needed to properly account for the effects of the strong magnetic fields that could be sustained by NS \cite{Lis2}. 

In terms of Eqs.~\eqref{Omega2} and \eqref{Nfree}, the pressure, energy density and heat capacity per unit of volume inside the star can be expressed as \cite{quintero2021}
\begin{equation}
	\begin{split}
P = - \Omega + \frac{1}{2} u_0 N^2_T,\qquad
E = - \dfrac{3}{2} \Omega + \frac{1}{2} u_0 N^2_T + m N_T,\\
C_V =  (2s+1)\left( \dfrac{m}{2\pi\beta}\right)^{\frac{3}{2}}\left[ \dfrac{15}{4}P_{5/2} (e^{\beta\mu'}) - \dfrac{9}{4} \dfrac{P_{3/2}^2 (e^{\beta\mu'})}{P_{1/2} (e^{\beta\mu'})}  \right]
\end{split}\label{EoS}
\end{equation}

\section{Macroscopic observables}\label{sec:structure equations}\label{sec3}

In this section, we discuss the system of equations we use to obtain the macroscopic properties of the stars and explain how to compute the mass-radius curves, the specific heat, and the tidal deformability. 

\subsection{Structure equations}

We start by considering a static, non-magnetized, isolated compact object described by the metric \citep{shapiro}
\begin{equation}
g^0_{\mu\nu} = - e^{\nu(r)}dt^2 + e^{\lambda(r)} dr^2 + r^2(d\theta^2 + \sin^2\theta d\phi^2);
\end{equation}
in spherical coordinates. Moreover, because in this case there exist no anisotropies, the energy-momentum tensor (EMT) of the matter composing the star is that of a static perfect fluid \citep{shapiro}
\begin{equation}
T^\mu_\nu = \textbf{diag} (-E(r), P(r), P(r), P(r)),
\end{equation}
where $E$ and $P$ are the pressure and energy density. For the BECS we are analyzing, they are related through Eqs.~\ref{EoS}. 

The metric and the EMT are inserted in Einstein's field equations $G^\mu_\nu = 8\pi G T^\mu_\nu$ \footnote{$G=6.711\times 10^{-45}$ MeV$^{-2}$ is the gravitational constant} to obtain the Tolman-Oppenheimer-Volkoff (TOV) equations \citep{shapiro}
\begin{equation}
\begin{split}
	e^{\lambda} &= 1 - 2\frac{Gm}{r} \qquad
	m' = 4 \pi r^2 E \\
	P' &= - \frac{1}{2} (E + P) \nu'(r)  \qquad
	\nu' = \frac{2G(m + 4\pi r^3 P)}{r(r - 2Gm)}.
\end{split}\label{TOV}
\end{equation}
Eqs.~\eqref{TOV} guarantee the hydrostatic equilibrium of the star in the context of general relativity and allow us to obtain its macroscopic properties, in particular, the mass  ($M$), radius ($R$) and specific heat (C) of the star. To do so we integrate Eqs.~\eqref{TOV} starting from the initial conditions $P(0)=P_c$ and $m(0)=0$ at the center of the star and ending at the value $R$ of the radial coordinate where the pressure vanishes, $P(R)=0$. This defines the radius of the star $R$, and its mass $M=m(R)$. To compute the heat capacity, we integrate the differential equation 
\begin{equation}
	\dfrac{dc_v}{dr}= \dfrac{4\pi r^2 C_V}{\sqrt{1-\dfrac{2Gm}{r}}}
\end{equation}
coupled to Eqs.~\eqref{TOV} and define the heat capacity of the star as $C=c_v(R)$. Also, the metric coefficient $\nu$ is normalized so that $\nu(R) = \left(1 - \dfrac{2GM}{R}\right)$.

\subsection{Tidal deformability}

To compute the tidal deformability let us consider that the static, spherically symmetric star of mass $M$ and radius $R$ of the previous section is placed in a static external quadrupolar tidal field $\epsilon_{ij}$. Such a star would depart from spherical symmetry and develop a quadrupolar mass moment $Q_{ij}$ ~\citep{Flores,Hinderer}. In an asymptotically mass-centered reference frame with cartesian coordinates, the metric function $g_{tt}$ can be expanded as ~\citep{Hinderer}
\begin{align}
	\dfrac{1 - g_{tt}}{2} &= - \dfrac{GM}{r} - \dfrac{3 Q_{ij}}{2r^3} \left(n^i n^j - \dfrac{1}{3}\delta^{ij}\right) \nonumber \\
	&+ \dfrac{1}{2}\epsilon_{ij} r^2 n^i n^j +o(r^3) + o\left(\frac{1}{r^4}\right)
	\label{general expansion}
\end{align}
where the Einstein summation convetion is assumed and $\vec{n} = (\sin\theta\cos\phi,\sin\theta\sin\phi,\cos\theta)$.  In the expansion Eq.~\eqref{general expansion}, the right-hand side is the Newtonian potential of the source (first two terms) plus that of the tidal field. To first order in $\epsilon_{ij}$, the tidal deformability parameter of the stars is defined as $\lambda = - Q_{ij}/\epsilon_{ij}$ \citep{Hinderer,Flores}. However, for practical reasons, we will use its dimensionless versions \footnote{Note that the names of these parameters vary in the literature.} : the second tidal Love number $k_2 = \frac{3\lambda}{2R^5}$ and the dimensionless tidal deformability $\Lambda = \lambda/(GM)^5$.

For axially symmetric perturbations, $Q_{ij}$ and $\epsilon_{ij}$ are diagonal traceless tensors that can be written in the form: $Q_{ij} =  Q \textbf{diag} (-1,-1,2)$ and $\epsilon_{ij} =  \epsilon \textbf{diag} (-1,-1,2)$. Then $\lambda = - Q/\epsilon$ and 
\begin{align}
	\dfrac{1 - g_{tt}}{2} &= - \dfrac{GM}{r} - \dfrac{3 Q}{r^3} P_2(\cos\theta) \nonumber \\
	&+\epsilon r^2 P_2(\cos\theta) +o(r^3) + o\left(\frac{1}{r^4}\right)
	\label{expansion}
\end{align}
where $P_2$ is the Legendre polynomial of order $2$.

Now, we can write the metric (the EMT) as a sum of the metric (the EMT) for the spherically symmetric space-time $ g^0_{\mu\nu}$($T^{\mu}_{\nu}$) and the tidal perturbation $h_{\mu\nu}$ ($\delta T^{\mu}_{\nu}$)
\begin{equation}\label{perturbation}
	\begin{split}
	h_{\mu\nu} &= \textbf{diag} ( -e^{\nu(r)} H_0(r), e^{\lambda(r)} H_2(r), r^2 k(r), r^2\sin^2\theta k(r)) \\
	&\times P_2(\cos\theta) , \\
	\delta T^\mu_\nu &=  \delta p(r,\theta) \textbf{diag}(-E'(P), 1,1,1).
	\end{split}
\end{equation}
With the help of Eq.~\eqref{perturbation} we linerarize Einstein equations $\delta G^\mu_\nu = 8\pi G \delta T^\mu_\nu$ and obtain for $H_0$ the following differential equation \citep{Hinderer, Flores}
\begin{equation}
	\begin{split}
	H_0'' &+ H_0' \left( \frac{2}{r} + \frac{\nu' - \lambda'}{2} \right) + H_0 \big[ - \frac{6}{r} e^{\lambda}  + \frac{3\lambda'}{2r}\nonumber \\
	& + \frac{7\nu'}{2r} + (\lambda'+\nu')\frac{E'(P)}{2r} - \frac{\lambda'\nu'}{2} - \frac{\nu'^2}{2} + \nu'' \big] = 0.
	\end{split}
\end{equation}\label{Ec H generica}

To obtain the perturbed metric outside the star, one imposes to Eq.~\eqref{Ec H generica} vacuum conditions: $m(r) = M$, $E(r)=P(r)=0$, $E'(r)=0$, and $e^{\nu(r)} = e^{-\lambda(r)} = \left(1 - \frac{2GM}{r} \right)$. With them, the exterior version of Eq.~\eqref{Ec H generica} takes the form
\begin{equation}
	\begin{split}
H_0'' + H_0'\left( \frac{2}{r} - \lambda' \right) - H_0 \left( \frac{6}{r^2}e^{\lambda(r)} + \lambda'^2 \right) = 0,
\end{split}
\end{equation}
which posseses an exact exact solution  \citep{Hinderer, Flores}:
\begin{equation}
H_0(r) = c_1 Q^2_2 \left( \frac{r}{GM} - 1\right) + c_2 P^2_2 \left( \frac{r}{GM} - 1\right),
\end{equation}
where $P^2_2$ and $Q^2_2$ are the $l=2,m=2$ associated Legendre functions of the first and second kind. 

Once the exterior expression of $g_{tt}$ is found, the expansion at the buffer zone is
\begin{equation}
	\begin{split}
	\dfrac{1 - g_{tt}}{2} &\approx - \frac{GM}{r} + \frac{4 c_1}{5}\left( \frac{GM}{r}\right)^3 P_2(\cos\theta) \\
	&+ \frac{3}{2} c_2 \left(\frac{r}{GM}\right)^2 P_2(\cos\theta) + \ldots
\end{split}
\end{equation}
from which the quadrupolar moments and the tidal deformability can be inferred as
\begin{equation}
	\begin{split}
Q = - \frac{4}{15} c_1 (GM)^5, \quad \epsilon=\frac{3 c_2}{2 (GM)^2},\quad \Lambda = \frac{8}{45}\frac{c_1}{c_2}.
\end{split}
\end{equation}
To compute the ratio $c_1/c_2$ the continuity of $H_0$ and $H_0'$ at the surface of the star is used \citep{Hinderer,Flores}, specifically the continuity of the function $y(r) = r\frac{H_0'(r)}{H_0(r)}$, that satisfies the differential equation 
\begin{equation}
\begin{split}
	r y'(r) &+ y'(r)^2 + F(r) y(r) + r^2 Q(r) = 0,\\
		F(r) &= \dfrac{1 - 4\pi G r^2 (E-P)}{1-2\frac{Gm}{r}}, \\
	Q(r) &= \dfrac{4\pi G}{1-2\frac{Gm}{r}} \left[ (P+E) E'(P) + 9P+5E - \dfrac{6}{4\pi Gr^2}\right], \\
	&- \dfrac{4G^2(M+4\pi r^3P)}{r^2(r-2GM)^2}, 
\end{split}
\end{equation}
which must be solved coupled with the TOV system with initial condition $y(0) = 2$. Once the value $y_R = y(R)$ is determined, the tidal deformability is computed as
\begin{equation}
	\begin{split}
	\Lambda &=  16 (1 - 2 C)^2 (2 + 2 C (-1 + yR) -  yR)  \\
	&\times \{30 C (6 - 3 y_R +  C (3 (-8 + 5 y_R)  \\ 
	&+ 2 C (13 - 11 y_R + C (-2 + 3 y_R + 2 C (1 + y_R))))) +  \\
	&45 (1 - 2 C)^2 (2 + 2 C (-1 + y_R) - y_R) \log(1 - 2 C)\}^{-1}.
\end{split}
\end{equation}
Let us note that this treatment doesn't take into account the deformation in the surface of the star and that the star's mass $M$ and radius $R$ remain the same as in the unperturbed configuration. Besides, the condition $|\delta p| << P$, needed for a perturbative solution, doesn't hold near the surface of the star. Nevertheless, here we follow this approach of \citep{Hinderer} and \citep{Flores} as a first approximation to compute the tidal deformability.

\section{Observational constraints}\label{sec4}

In the following subsections we evaluate the BECS model in the light of three recent observational constraints: the measurement of the lightest neutron star known so far, the central CS in the supernova remnant HESS J1731-347; the gravitational wave event GW170817; and the new lower bound found for the heat capacity of the neutron stars core. 

\subsection{Strangely light NS as a Bose-Einstein condesate star}\label{LS}

In \citep{Doroshenko2022} the authors reported a mass of $M=0.77^{+0.2}_{-0.17}M_\odot$ corresponding to a radius of $R=10.4^{+0.86}_{-0.78}$ km for the compact object XMMU J173203.3-344518 (COO from now on) within the supernova remnant HESS J1731-347. This is the lightest NS known so far, and possible explanations for its low mass include, for instance, that the COO is a strange star \citep{Doroshenko2022, Horvath2023}, or a NS that contains a non-negligible amount of dark matter \citep{Diedrichs2023}. Here, we show that given the mass and radius of COO, it could be also a NS with an interior of condensed paired neutrons, i.e., a BECS. 

\begin{figure}[h]
	\begin{center}
		\includegraphics[width=.8\linewidth]{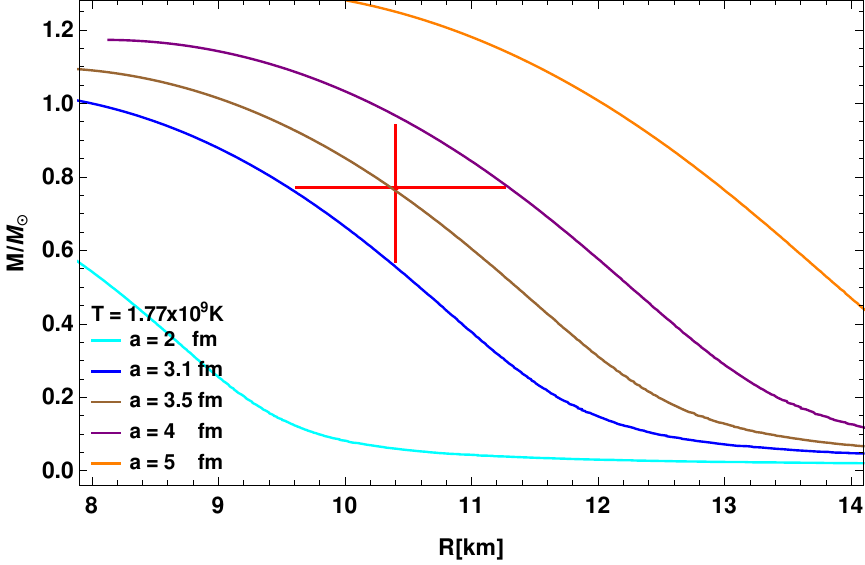}
	\end{center}
	\caption{Mass-radius relation for BECS with scattering length from $2$ to $5$ fm and $T=1.77\times 10^9$ K. The error bars correspond to the CS within the supernova remnant HESS J1731-347 \cite{Doroshenko2022}.}
	\label{M vs R}
\end{figure}

In Fig. \eqref{M vs R} we draw the mass-radius relation for BECS setting the scattering length $a$ from $2$ to $5$ fm, along with the result of the observation of the mass and radius of the COO previously mentioned. The temperature of BECS has been fixed to $T=1.77\times 10^9$ K since this value was reported for the COO in \citep{Horvath2023}.   
Note that the mass and radius of BECS increase with $a$, and that curves with $3.1 \leq a \leq 4$~fm, intersect
the region inside the error bars for the estimated parameters of
the CCO. Therefore, we suggest that this strangely light neutron star could be a BECS with a scattering length approximately between $3.1$ and $4$~fm.

\subsection{Constraints from tidal deformability}\label{TD}

The only measurement of the tidal deformability of NS is reported in \citep{Abbott2019} and it is based on the data coming from the compact binary inspiral event GW170817. The combined tidal deformability of the NS in this event is 
\begin{equation}\label{lambda}
\Lambda_c = \dfrac{16}{13} \dfrac{ (m_1 + 12m_2)m_1^4 \Lambda_1 + (m_2+12m_1) m_2^4 \Lambda_2}{(m_1+m_2)^5} = 300^{+420}_{-230},
\end{equation}
where $\Lambda_1$ and $\Lambda_2$ are the tidal deformabilities of the component stars, and $m_1$ and $m_2$ the corresponding masses, which are estimated to be in the range $1.16-1.60 M_\odot$ \citep{Abbott2019}. 

\begin{figure}[h]
	\begin{center}
		\includegraphics[width=.8\linewidth]{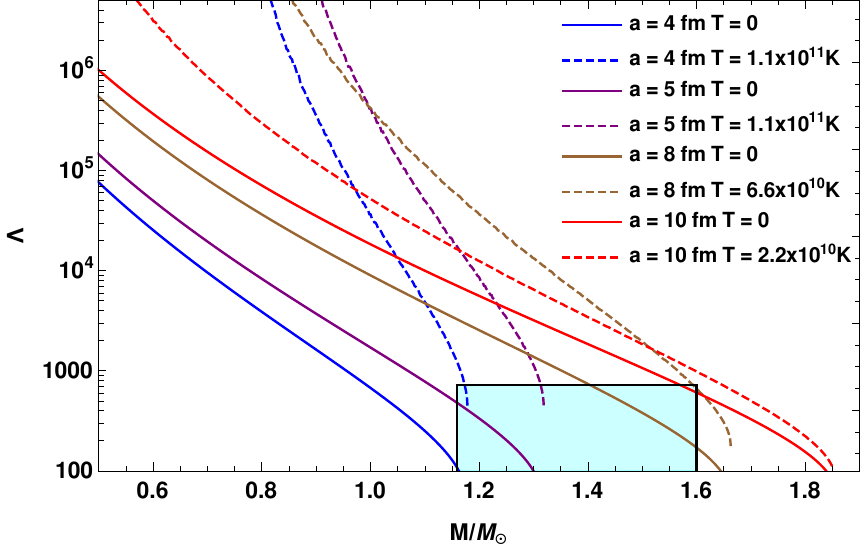}
	\end{center}
	\caption{Tidal deformability of BECS as a function of the mass of the stars for several values of scattering length and temperature. The colored box signals the bounds inferred from the GW170817 event\cite{Abbott2019}.}
	\label{Lambda plot}
\end{figure}

Since $\Lambda_c = \Lambda_1 = \Lambda_2$ if $m_1=m_2$, we use its upper bound, $\Lambda_{max} = 720$, as an upper bound to $\Lambda_{1,2}$. Fig. \eqref{Lambda plot} shows the relation $\Lambda(m)$, computed with the EoS for BECS. We used several values of $a$ and $T$ to constrain these magnitudes using the bounds in mass and tidal deformability from GW170817. 

Let us first note that both, the tidal deformability and the mass of the BECS grow with $a$ and $T$. Moreover, for a fixed mass of the star, a higher $a$ or $T$ implies a higher tidal deformability. This expresses the fact that the higher the temperature or the stronger the repulsive interaction, the less compact and more deformable the star \cite{Lis2}. Varying both magnitudes it is possible to find BECSs that intersect the colored box that signals the bounds inferred from the GW170817 event in Fig. \eqref{Lambda plot}. In this case, we obtain that the BECS model for neutron stars is compatible with the tidal deformability estimated for the stars in the  GW170817 event, for scattering lengths approximately between $4$ and $10$~fm, and temperatures up to $10^{10}$~K.

\subsection{Lower bound of the heat capacity in neutron stars cores}\label{HC}

Finally, we evaluate the BECS as a model for NS cores in the light of the results presented in \citep{Cumming2017}. There, the authors carried out a study of observational data of the thermal evolution of several NS and, using different EoS models for the NS envelope, they established a lower bound for the heat capacity of the core: $C_0 = 1.4-5.0 \times 10^{52}\frac{T}{10^8K}$.
\begin{figure}[ht]
	\begin{center}
		\includegraphics[width=.8\linewidth]{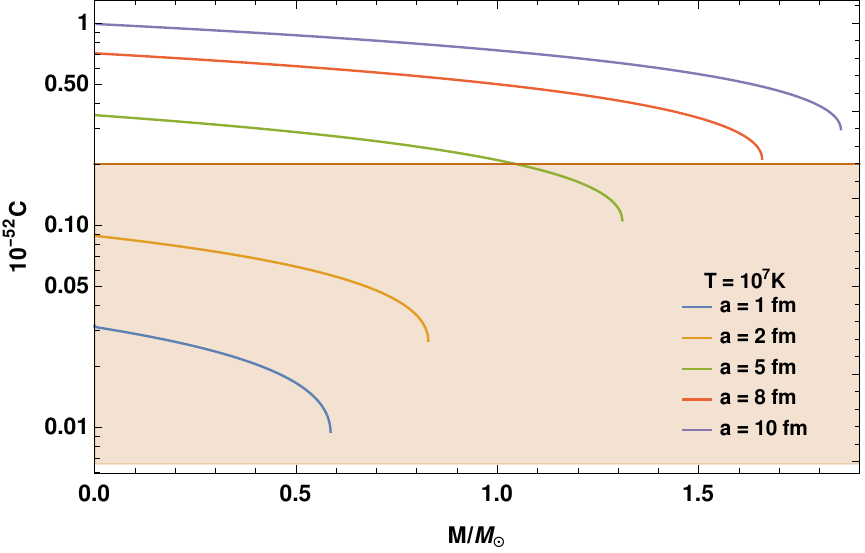} 
	\end{center}
	
	\begin{center}
		\includegraphics[width=.8\linewidth]{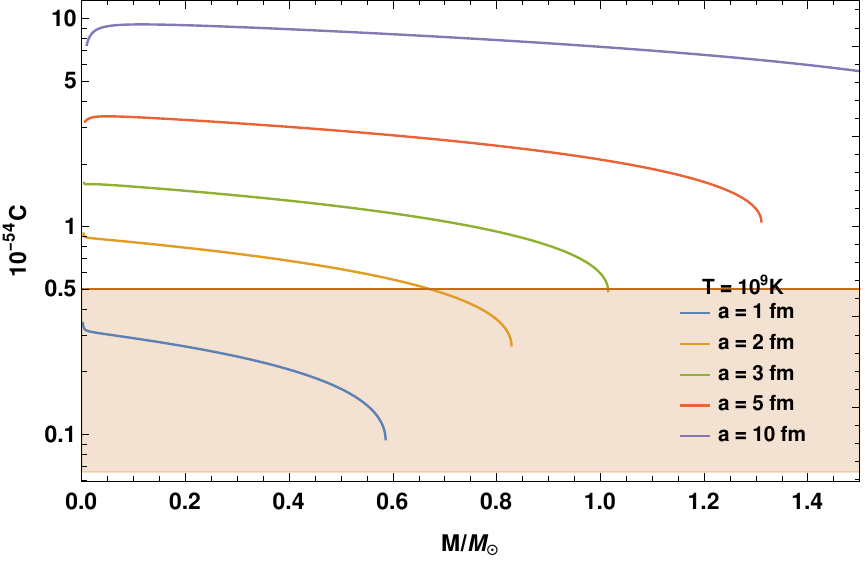} \qquad
	\end{center}
	\caption{The heat capacity of BECS as a function of the mass of the star for several values of the scattering length. The upper panel corresponds to  $T=10^7$ K and the lower panel, to $10^9$ K. The horizontal lines mark the corresponding lower bounds.}
	\label{C vs M}
\end{figure}
Fig.~\eqref{C vs M} shows the heat capacity of BECS as a function of the mass of the star for several values of $a$ and temperatures $10^7$ K and $10^9$ K. The values of $a$ needed to fulfill the heat capacity constraint decrease as the temperature increases. Restricting the analysis to masses higher than $0.7$ M$_\odot$ we found that, if $T=10^7$ K, then we need $a\ge 5$ fm, while for $T=10^9$ K, then $a\ge 2$ fm. This means that, for fixed $a$ and $M$, there exists a lower bound in the temperature a BECS can have. It is also worth noticing that for $T=10^9$~K, this result is compatible with that of subsection \ref{LS}, where the observational data of a light NS with a temperature in the same order was considered. Hence, in general, we can affirm that BECS as models for NS cores are also able to fulfill the specific heat constraint.

\section{Conclusions}\label{sec5}

We have used up-to-date observational data to evaluate BECS as a model for NS. In Section \eqref{TD} we constrained the scattering length of naked BECS to $4$ fm $\le a \le 10$ fm for temperatures $T\le 10^{10}$ K from the mass and tidal deformabilities of the stars involved in GW170817. Results of Section \eqref{LS} suggest the inclusion of naked BECS with $3.1$fm $\le a \le 4$fm and $T\sim 10^{9}$ K in the list of exotic NS models that could explain the nature of the ultralight compact object XMMU J173203.3-344518. Finally, we found that to fulfill the lower bound of the core of heat capacity of NS, BECSs with $T\sim 10^7$ K require $a\ge 5$ fm, while BECSs with $T\sim 10^9$ K need $a\ge 2$ fm. This last range is consistent with the estimated range for $a$ that comes from Section \eqref{LS}. Moreover, if $a$ were known from experiments or theory, one might be able to set a mass-dependent lower bound in the BECS core temperature.

So far our theoretical treatment of BECS does not include several features of NS that may have an important influence on the macroscopic properties, such as the star rotation, its magnetic field, and the presence of the crust and the atmosphere. For instance, rotation and magnetic field deform the star's surface and increase the mass and radius of NS sequences compared to their static and non-magnetized counterparts with the same central density. Also, the magnetic field effects on the heat capacity of a system of vector bosons are important, as demonstrated in \cite{quintero2021}. Therefore,  although the predictions derived from naked BECS as NS models with approximately  $a\sim 4$ fm fits all the observational data here considered, an improvement of this estimate may come in future investigations from new observational data and the inclusion of rotation, magnetic field, and the crust and atmosphere in the BECS models.

\section*{Supporting information}

This work was supported by Project No. NA211LH500-002 of AENTA-CITMA, Cuba. 	G. Quintero Angulo gratefully acknowledges the support of the Alexander von Humboldt Foundation.

\bibliography{Magic_2023_v03}

\end{document}